\documentclass[a4paper]{jpconf}
\usepackage{graphicx}
\usepackage{siunitx}
\usepackage{url}
\usepackage[hidelinks]{hyperref}
\begin{document}
\title{Study of the transfer and matching line for a PWFA-driven FEL}
\author{P Iovine$^{1,2}$, A Bacci$^4$, 
A Biagioni$^{3}$, E Chiadroni$^{1,3}$, L Crincoli$^{3}$, A Del Dotto $^{3}$, M Ferrario$^{3}$, A Giribono$^{3}$, R Pompili$^{3}$, S Romeo$^{3}$, M Rossetti Conti$^4$ and C Vaccarezza$^3$}
\address{$^1$ Sapienza, University of Rome, 00161, Rome, Italy}
\address{$^2$ INFN, Sezione di Napoli, 80126, Napoli, Italy}
\address{$^3$ INFN, Laboratori Nazionali di Frascati, 00044, Frascati, Rome, Italy}
\address{$^4$ INFN, Sezione di Milano, via Celoria, 16, 20133, Milano, Italy}
%\ead{pasqualina.iovine@uniroma1.it}
\begin{abstract}
 The development of compact accelerator facilities providing high-brightness beams is one of the most challenging tasks in the field of next-generation compact and cost affordable particle accelerators. Recent results obtained at SPARC\_LAB show evidence of the FEL laser by a compact (3 cm) particle driven plasma-based accelerator. This work is carried out in the framework of the SPARC\_LAB activities concerning the R$\&$D on plasma wakefield accelerators for the realization of new compact plasma based facilities, i.e EuPRAXIA@SPARC\_LAB. The work here presented is a theoretical study demonstrating a possible scheme concerning the implementation of an innovative array of  discharge capillaries, operating as active-plasma lenses, and one collimator to build an unconventional  transport line for bunches outgoing from plasma accelerating module. Taking advantage of the symmetric and linear focusing provided by an active-plasma lens, the witness is captured and transported along the array without affecting its quality at the exit of the plasma module. At the same time the driver, being over-focused in the same array, can be removed by means of a collimator.
\end{abstract}

\section{Introduction}
Compact accelerator facilities development, providing high-brightness beams, is one of the most challenging tasks in the field of next generation compact and cost affordable particle accelerators. Particular attention is given to plasma-driven particle facilities due to their possibility to integrate high-gradient accelerating plasma modules with a short wavelength Free Electron Laser (FEL). Extremely strong electric fields (up to hundreds of GV/m) generated in the plasma allow accelerating gradients much higher than in conventional accelerators and set the basis for achieving very high final energies in a compact space.
 The plasma-based accelerator scheme is considered an efficient alternative to the current RF accelerators capable of ensuring compact structures \cite{cakir2019brief}. \\The  EuPRAXIA (European Plasma Research Accelerator with eXcellence In Applications) preparatory phase aims at designing of the world’s first accelerator facility based on advanced plasma-wakefield techniques to deliver 1-5 GeV  high brigthness beams as required for users applications. This project foresees the realisation of a user facility  based on both laser-driven and beam-driven plasma acceleration. For the beam-driven scenario the LNF-INFN laboratories in Frascati (Italy), with its EuPRAXIA@SPARC\_LAB projects, represents one of the pillars for the experimental activities. Indeed, at SPARC\_LAB, recent experimental results represented a first proof-of-principle demonstration of FEL lasing by a compact (3 cm) particle beam-driven plasma accelerator \cite{pompili2021first, galletti}.
\section{Motivation}
The work presented here is in the framework of the EuPRAXIA@SPARC\_LAB project \cite{assmann2020eupraxia} and aims to demonstrate the feasibility of realising a transport line for beams outgoing from a plasma module preserving beam parameters.
\\In plasma-based acceleration, either driven by ultra-short laser pulses \cite{tajima1979laser} or electron bunches \cite{litos2014high} the plasma is used as an energy transformer in which the driver pulse energy is transferred to the plasma through the excitation of plasma wakes and, in turn, to a witness bunch externally \cite{rossi2014external} or self-injected \cite{geddes2004high}.
A common issue to both the laser and particle beam-driven methods is represented by the extraction of the accelerated witness bunch that can lead to a large degradation of its emittance.\\ When exiting the plasma the accelerated bunch has a large angular divergence, of several mrad, that is some orders of magnitude larger with respect to beams accelerated by conventional (RF) photo-injectors. For non-negligible energy spreads $\sigma_E$, such a large divergence also leads to a rapid increase of the normalized emittance that, in a drift of length $s$, is given by the following formula:
\begin{equation} 
    \varepsilon ^2 _n=  \bigl\langle  \gamma \bigr\rangle ^2 \Biggl( s^2 \bigl(\frac{\sigma_E}{E}\bigr)^2 \sigma^4_{x'} + \varepsilon^2_g \Biggl)
\end{equation}\label{emigrowth}
where $E$ is the bunch energy and $\varepsilon_g$ the geometrical emittance \cite{migliorati2013intrinsic}.
 Thus, it is mandatory to catch the accelerated witness as soon as possible. In addition because of the high energy spread, the electron bunch coming from plasma accelerators suffers of strong chromatic effects that quickly degrade the emittance. An analytical description of the degradation of beam emittance in free space propagation is given by the chromatic length parameter $L_C = \frac{\gamma(\sigma^0_x)^2}{\sigma_E \varepsilon_0}$ where $\gamma$ is the Lorentz factor, $\sigma_E$ the relative energy spread, $\sigma^0_x $ and $\varepsilon_0$ are the initial values of the transverse size of the bunch and geometric emittance. $L_c$ shows how normalized emittance dilution is driven both by a high energy spread and a small ratio between the beam size and divergence \cite{conti2018electron}. 

\section{EXTRACTION SYSTEM}
In this paper, an innovative extraction system based on active plasma lenses will be presented. Two active-plasma lenses (APLs) \cite{van2015active} capture and focus the witness while a collimator is used to remove the driver.
An APL essentially consists of a cylindrical current-carrying conductor whose axis is parallel to the beam. Here, the plasma, produced after the ionization of the gas confined within the capillary, only acts as a conductor while the net focusing effect is produced by the flowing discharge current. In fact, according to Ampere's law, the bunch is focused by the azimuthal magnetic field, $B_\phi$ generated by the discharge current:
\begin{equation}
  B_\phi(r)= \frac{\mu_0}{r} \int_{0}^{r} J(r') r' dr' 
\end{equation}\label{ampere1} 
where $\mu_0$ is the vacuum permeability and $J(r')$ the current density within the aperture (r$<$ R with R the capillary radius). If we consider the focusing strength of an APL, given by
$ K= \frac{\partial B_\phi (r)}{\partial r} \frac{e_0}{m_0 c \gamma}$, it can be understood how these elements are innovative. In fact, their focus is symmetric like in solenoids, but the resulting force scales as F $\propto$ $\gamma^{-1}$ (with $\gamma$ the relativistic Lorentz factor) like in quadrupoles. Moreover, the focusing can reach several tens of kT/m, that is, orders of magnitude larger than the strongest available quadrupoles ($\simeq$ 600 T/m). The tunability of the system is obtained by adjusting the external discharge current $I_D= \int_S \textbf{J} \cdot· d\textbf{S}$.\\Driver removal is based on the different focusing provided by the first APL to the accelerated witness bunch and the energy-depleted driver: the witness is focused exactly at the entrance of a collimator so as not to cut its charge while the driver is over-focused to have a spot size larger than the collimator aperture. %%%%%%%%

Starting from a recent work \cite{pompili2019plasma}, the simulation analysis, which led to the definitive choice of the characteristics of the elements along the line, was conducted by using several simulation tools linked each other. The plasma acceleration process was simulated with Architect code, a hybrid code that works as a Particle-In-Cell (PIC) for the electron bunches while treating the plasma as a fluid \cite{marocchino2016efficient}. With Architect code a driver of 170 pC followed by a witness of about 30 pC are simulated downstream the PWFA module. Witness and driver bunches parameters at the exit of the accelerating module are in the Table \ref{tab1}.
\begin{table}[h]
   \caption{ \label{tab1}Bunches parameters at the exit of the PWFA module. {$^\ast$ Witness core parameters are rms value referring to the 80\% of the whole bunch.}}
   \begin{center}
\begin{tabular}{lllll}
\br
       \textbf{Parameter} & \textbf{Units} & \textbf{Witness} & \textbf{Witness core}$^\ast$ & \textbf{Driver} \\
       \mr
         Charge & pC &30&24&169\\
Duration (rms) &fs &18 &7&160\\
Energy& MeV &1000&990 &300\\
Energy Spread (rms)&\% &1.20&0.06& 56.86\\
Emittance & $\si{\micro\meter} $&0.5& 0.4&2.6\\
Spot size &$\si{\micro\meter} $&1& 1&4\\ 
\br

\end{tabular}

\end{center}

\end{table}
The dynamics of the two bunches at the exit of the PWFA module is simulated by General Particle Tracer (GPT) code, while MATLAB (MATrix LABoratory) is used for the analysis of beam propagation in plasma lenses. The first code takes into account space charge effects, while the second code reproduces the magnetic field generated by the discharge current in APL. Moreover, MATLAB code is used to simulate beam interaction with the plasma considering plasma wakefields acting on the beam itself using the equation reported in \cite{fang2014effect}. 
Thanks to the use of these tools, a possible scheme concerning the implementation of an array of active plasma lenses and a collimator was simulated to design a transfer line capable of extracting and transporting the accelerated and highly divergent witness bunch and, at the same time, to reduce the driver to 1.4\% of its initial charge. 
The loss of the witness bunch is negligible, amounting to only 0.07\% of its initial charge.

\begin{figure}[h]
\begin{center}
    \includegraphics[scale=0.7]{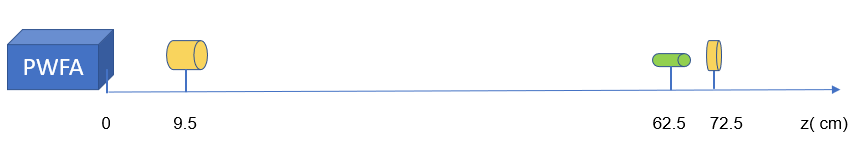}
   \end{center}
 \caption{\label{figura:2}Layout, not in scale, of the extraction system consisting of two active-plasma lenses (in yellow) and a collimator between them (in green).}
\end{figure}

\begin{figure}
\centering
\includegraphics[width=20pc]{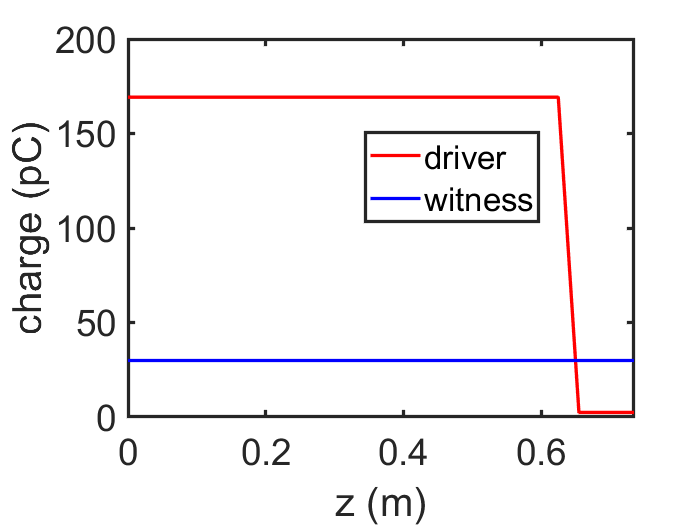}%

 \caption{ \label{fig:carica} Witness (in blue) and Driver (in red) charge degradation along the beam line.}

\end{figure}

Figure ~\ref{figura:2} shows the layout of the extraction system, while the optimized parameters are listed in Table \ref{tab2}. 
\begin{table}[h]
   \centering
   \caption{ \label{tab2}Optimized parameters for the lenses and collimator used in the extraction system. The position of each element is relative to the exit of the PWFA module.}
   \begin{center}
\begin{tabular}{lllll}
\br
       \textbf{Element} & \textbf{Length (cm)}  &\textbf{Radius ($\si{\micro\meter} $)}& \textbf{Position (cm)} &\textbf{Current (kA)}\\ 
    
       \mr
 Lens1     & 3 & 500 & 9.5 & 1.0\\
Collimator & 3 & 150 & 62.5& \\
Lens2      & 1 & 500 & 72.5 &0.4\\
       \br
   \end{tabular}
\end{center}
\end{table}
The lenses are 3 cm and 1 cm, respectively long and both have a 500 $\si{\micro\meter}$ hole radius.
The focusing is obtained applying a discharge current $I_D$ of 1 kA in the first and 400 A in the second lens. The position of the first lens with respect to the PWFA module has been carefully chosen to preserve the witness normalized emittance taking into account different contributions and seeking a compromise between them. In fact, as shown in Eq.1, %\ref{emigrowth}
short drifts are preferable to prevent emittance degradation due to the large divergence of the beam. At the same time, small drifts  lead to a smaller witness spot at the APL entrance, resulting in a higher bunch density. As demonstrated a few years ago in \cite{pompili2} high bunch densities produce non-negligible plasma wakefields in the APL that, being nonlinear, would increase the beam emittance during the lens focusing. The current of the first lens is higher than in the second to provide stronger focusing  in the first centimeters of the line and to cut more driver at the entrance of the second lens, which in this configuration also acts as a collimator.\\Taking advantage of the focusing of the first lens, a collimator 3 cm long with an aperture of 150 $\si{\micro\meter}$ is located at 62.5 cm from the accelerating module. At this position the witness is focused exactly at the entrance of the collimator, without losing any charge, while the driver is cut because it is overfocused to a spot size much larger than the collimator aperture almost in its entirety (see Fig. \ref{fig:carica}).

\begin{figure}[h]
\begin{minipage}{19pc}
  % \begin{center}
   \includegraphics[width=19pc]{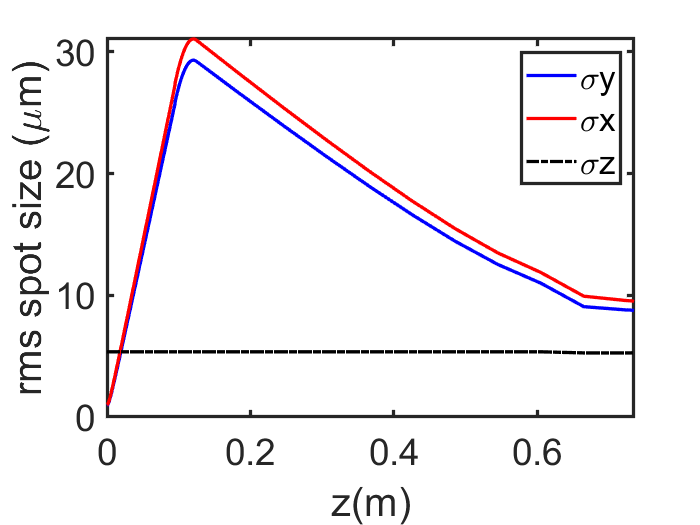}
   %\end{center}
   \caption{\label{fig:3sigma}Witness envelope along the beam line.}
  \end{minipage}\hspace{1.5pc}%
\begin{minipage}{18pc} 
%\end{figure}

%\begin{figure}[h]
%\begin{center}
   \includegraphics[width=19pc]{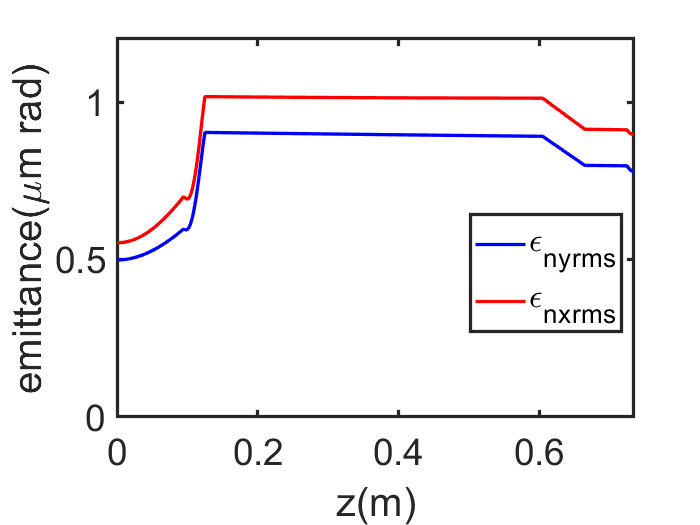}
  % \end{center}
   \caption{\label{fig:3emi} Evolution of the witness normalized emittance along the beam line.}
  \end{minipage} 
\end{figure}

From parametric studies performed by varying the element settings, this configuration appears to be a good compromise to guarantee transport of the witness in a very compact length (73.5cm), as highlighted in Fig. \ref{fig:3sigma}.
Figure \ref{fig:3emi} shows that witness normalized emittance along beamline grows from 0.5 mm mrad to 0.8 mm mrad. As expected, this growth occurs primarily in the initial drift downstream of the PWFA module and within the APLs. 
The changes observed in the beam envelope and emittance trend, respectively in Fig. \ref{fig:3sigma} and Fig. \ref{fig:3emi}, are attributed to the 0.07\% charge degradation of the witness after the collimator.

%\end{figure}

 Such a solution is tunable by adjusting the parameters of the elements along the line and can be implemented in a future facility based on plasma acceleration, where the compactness represents the main goal. 
\section{CONCLUSION}
Plasma-based acceleration,  driven either by ultra-short laser pulses or electron bunches, has proven the feasibility to drive a high-gain FEL. The focus of the research now concerns on one side on the reproducibility and stability of the acceleration, mandatory to pilot a user-facility and, on the other side, the proper extraction and capture of the accelerated beam to avoid its quality degradation, which might prevent FEL generation. \\
This work proposes a possible scheme concerning the implementation of an innovative array of discharge capillaries, operating as active-plasma lenses, to build an unconventional transport line for bunches outgoing from plasma accelerating module.
Taking advantage of the symmetric and linear focusing provided by an active-plasma lens and a use of a collimator, this compact and tunable configuration appears to be a good compromise to guarantee transport of the witness and the driver removal in a 73.5 cm length. This beam line involves a 60\% increase in witness emittance compared to the initial value but, despite this increases, it represents a valid alternative to  a conventional transfer line. Indeed, dogleg or chicane beam lines requires longer spaces (of the order of tens of meters) and, due to a finite energy spread, might strongly affect the beam longitudinal phase space, requiring small deflection angles and non-trivial methods to limit and compensate the bunch elongation due to the dispersion introduced by each bending.
Parametric studies are on going to show the system tunability by varying the energy and the energy spread of the beams. In addition, the system can be tuned by varying the current and position of the lenses.\\ The proposed solution represents a good compromise to guarantee transport of the witness and the driver removal in a very short length that could be implemented in the future plasma acceleration facility.
\ack
This work was supported by the European Union’s Horizon 2020 research and innovation programme under grant agreement No. 101079773 (EuPRAXIA) and by the 5th National Scientific Committee of the INFN with the SL COMB2FEL experiment.

\section*{References}
\bibliography{iopart-num}
\end{document}